\documentclass[12pt]{article}
\usepackage{pdproc,graphicx} 

\begin{document}

\renewcommand{\thefootnote}{\alph{footnote}}
  
\title{BEHIND PVLAS\footnote{Talk given at the ``XII International Workshop on Neutrino Telescopes'' (to appear in the Proceedings).} }

\author{MARCO RONCADELLI}

\address{INFN and Dipartimento di Fisica Nucleare e Teorica, Universit\`a di Pavia, \\ 
Via A. Bassi 6, I-27100 Pavia, Italy\\
{\rm E-mail: marco.roncadelli@pv.infn.it}}

\normalsize\baselineskip=15pt

\abstract{We offer a detailed review of the physics behind the PVLAS experiment. We also 
address some laboratory and astrophysical cross-checks for the recent claim concerning a signal consistent with an Axion-Like Particle. Our aim is that the leading role played by Emilio (Mimmo) Zavattini in this field of research will become apparent.} 

\normalsize\baselineskip=15pt

\section{Introduction}

What an experimental physicist would normally do in trying to check subtle predictions of QED would be to submit his proposal to the committee of an high-energy accelerator, hoping for approval. And this would be even more natural for an experimentalist looking for a new elementary particle. 

Emilio (Mimmo) Zavattini~\footnote{Deceased January 9, 2007.} took a totally unconventional attitude, in line with his original, independent and creative mind. In the first place, he soon realized that a high-precision laser experiment involving classical optics has a better chance to detect the QED vacuum polarization effects produced by a strong magnetic field. This was pointed out in 1979 in collaboration with Enrico Iacopini~\cite{Iacopini}. 

In the meantime, the axion~\cite{WW}  appeared on the scene of elementary-particle theory as a necessary consequence of the Peccei-Quinn mechanism~\cite{Peccei} proposed as a natural solution to the ``strong CP problem''~\cite{SCPP}. In 1986, Luciano Maiani, Roberto Petronzio and Mimmo~\cite{Maiani} recognized that just the same kind of laser experiment can lead to the discovery of the axion, thanks to its coupling to photons. Besides uncovering its existence, careful optics measurements of this sort can yield both the axion mass and its photon coupling. Moreover, interesting regions of the parameter space of Axion-Like Particles~\footnote{A precise definition of Axion-Like Particle will be given below. For the moment, it can be understood as a particle similar to the axion, however with no fixed relationship between mass and photon coupling.} can be probed by the same strategy. 

A big effort was devoted by Mimmo and his group to turn these intuitions into a real experiment. 

A first step along this avenue was taken with the Brookhaven experiment in 1992, whose negative result produced an upper bound on the axion-photon coupling~\cite{Brookhaven}. 

A more sophisticated experiment -- called PVLAS (Polarization of Vacuum with a LASer) -- is still operating at the Legnaro National Laboratory of INFN. In 2005, the PVLAS collaboration reported positive evidence for a signal consistent with an Axion-Like Particle~\cite{pvlas1}, and this finding was announced by Mimmo at a Workshop of this series~\cite{pvlas2}. Whether new physics has been discovered by the PVLAS experiment is not yet clear, and this point is discussed in the talk of Guido Zavattini at this Workshop~\cite{GZ}.

Our aim is to review the physics behind the PVLAS experiment, along with some laboratory and astrophysical cross-checks for the recent claim. We hope that the leading role played by Mimmo in this exciting area of research will become evident.

\section{QED vacuum effects in a magnetic field}

We shall be concerned throughout with a monochromatic photon beam with frequency $\omega$ propagating in vacuo along the $z$-direction, in a magnetic field ${\bf B}$. Indeed, the presence of an external ${\bf B}$ field is common to all considerations to follow. We suppose that ${\bf B}$ is homogeneous and lies at a nonvanishing angle $\theta$ with the wave vector ${\bf k}$ of the beam. In addition, we assume $B \ll B_{\rm cr}$, where $B_{\rm cr} \equiv m_e^2/e \simeq 4.41 \cdot 10^{13} \, {\rm G}$ denotes the {\it critical} magnetic field ($m_e$ is the electron mass).

Classically, the magnetized vacuum is a non-dispersive medium because of the linear structure of Maxwell equations. Therefore, its index of refraction is trivial and so $\omega = k$, which entails in turn that the properties of the beam are unaffected by ${\bf B}$ (we take 
$\hbar=c=1$ as usual).

Quantum corrections change the situation, since virtual fermion exchange produces an effective photon-photon interaction, thereby making the theory non-linear. At one loop, this comes about through a box diagram with four external propagating photon lines and internal fermion lines. In the approximation $\omega \ll m_e$, this photon-photon interaction is represented by the 
Heisenberg-Euler-Weisskopf (HEW) effective lagrangian~\cite{Heisenberg}
\begin{equation}
\label{a1} 
{\cal L}_{\rm HEW} = \frac{{\alpha}^2}{90 \, m^4_e} \, \left[ \left(F_{\mu \nu} \, F^{\mu \nu} 
\right)^2 + \frac{7}{4} \left(F_{\mu \nu} \, \tilde F^{\mu \nu} \right)^2 \right] = \frac{2 {\alpha}^2}{45 \, m^4_e} \, \left[ \left({\bf E}^2 - {\bf B}^2 \right)^2 + 7 
\left({\bf E} \cdot {\bf B} \right)^2 \right]~,  
\end{equation}
where $\alpha$ is the fine-structure constant, $F_{\mu \nu} = ({\bf E}, {\bf B})$ is the electromagnetic field strength, $\tilde F^{\mu \nu}$ is its dual and natural Lorentz-Heaviside units are employed throughout. It goes without saying that photon-photon interaction implies the existence of photon-photon scattering
\begin{equation}
\label{aa1}
\gamma + \gamma \to \gamma + \gamma~,  
\end{equation}
whose amplitude depends on the {\it polarization} state of the incoming photons and can be computed by means of lagrangian (\ref{a1})~\footnote{As already stated, the HEW effective lagrangian makes sense for $\omega \ll m_e$ only. We shall always suppose that this condition is satisfied.}. 

Very interesting phenomena arise when a magnetic field ${\bf B}$ is present. Symbolically, they emerge from eq. (\ref{aa1}) by replacing one or more propagating photons $\gamma$ by a ``magnetic-field photon'' ${\gamma}_B$. Just as before, the corresponding amplitude can be computed from the HEW effective lagrangian. By solving the resulting field equations, one discovers that the propagation eigenstates are photons with {\it linear} polarization, either {\it parallel} or {\it normal} to the plane defined by the vectors $\bf B$ and $\bf k$: these modes will be denoted by ${\gamma}_{\parallel}$ and ${\gamma}_{\bot}$, respectively~\footnote{We follow the convention to define photon polarization as the direction of the {\it electric} field (surprisingly, no general consensus exists on this point).}. 

We begin by performing the replacement $\gamma \to {\gamma}_B$ in both sides of eq. (\ref{aa1}). In this way, we get the process
\begin{equation}
\label{ab1}
\gamma + {\gamma}_B \to \gamma + {\gamma}_B~,  
\end{equation}
which is {\it Delbr\"uck scattering}~\cite{Delbruck}, namely photon scattering in a magnetic field. We recall in this connection that Furry's theorem~\cite{Furry} tells that diagrams with an odd number of photon vertices vanish. So, the triangle diagram with one ${\gamma}_B$ line vanishes, and the leading contribution to the photon propagator is presently given by the above box diagram with two ${\gamma}_B$ lines, which yields precisely process (\ref{ab1}).  Clearly, its amplitude is of order $\alpha^2 B^2$, but one factor of $\alpha$ can be absorbed into $B_{\rm cr}^2$ so that it actually goes like $\alpha \, (B/B_{\rm cr})^2$. Moreover -- as for process (\ref{aa1}) -- this amplitude depends on the polarization state of the incoming photon. As a consequence, the modes ${\gamma}_{\parallel}$ and ${\gamma}_{\bot}$ propagate with {\it different} velocities. In other words, the refractive indices of the two propagating modes $n_{\parallel}$ and $n_{\bot}$ are different, 
with~\cite{Dittrich}
\begin{equation}
\label{a2}
n_{\parallel} = 1 + \frac{7}{2} \, \left( \frac{\alpha}{45 \pi} \right) \, 
\left( \frac{B \, {\rm sin} \, \theta}{B_{\rm cr}} \right)^2~,
\end{equation}
\begin{equation}
\label{a3}
n_{\bot} = 1 + \frac{4}{2} \, \left( \frac{\alpha}{45 \pi} \right) \, 
\left( \frac{B \, {\rm sin} \, \theta}{B_{\rm cr}} \right)^2~.
\end{equation}
Thus, we see that the QED magnetized vacuum produces a selective change in the velocity of light depending on its polarization state, a phenomenon called {\it birefringence} in analogy with what happens in an anisotropic optical medium~\footnote{However -- at variance with the case of a material body like a crystal -- here birefringence is achromatic, since $n_{\parallel}$ and $n_{\bot}$ are independent of $\omega$.}.

Next, we perform the replacement $\gamma \to {\gamma}_B$ only in the l.h.s. of eq. (\ref{aa1}). Accordingly, we are lead to the process
\begin{equation}
\label{ac1}
\gamma + {\gamma}_B \to \gamma + {\gamma}~,  
\end{equation}
which represents {\it photon splitting} as described by the above box diagram with a single ${\gamma}_B$ line. However, a careful analysis shows that CP-invariance forces this diagram to vanish in the limit of no vacuum dispersion (namely for $\omega = k$)~\cite{Adler}. Of course, the previous analysis has shown that ${\bf B}$ does make the vacuum dispersive, but -- as it generally happens in the presence of a selection rule -- the naive estimate of the orders of magnitude involved would give a wrong answer. Specifically, the leading contribution to the photon splitting amplitude turns out to come from the {\it exagon} diagram with three 
${\gamma}_B$ lines. Symbolically, we have
\begin{equation}
\label{ad1}
\gamma + {\gamma}_B + {\gamma}_B + {\gamma}_B \to \gamma + {\gamma}~,  
\end{equation}
and so its amplitude goes like ${\alpha}^{3/2} \, (B/B_{\rm cr})^3$, which is by a factor $(B/B_{\rm cr})^2$ {\it smaller} than naively expected. Also in this case the amplitude depends on the polarization state of the incoming photon and an explicit calculation~\cite{Adler} shows that it {\it vanishes} for incoming ${\gamma}_{\parallel}$ photons. As a consequence, ${\gamma}_{\parallel}$ photons do not split. On the other hand, incoming ${\gamma}_{\bot}$ photons split predominantly into ${\gamma}_{\parallel}$ photons~\cite{Adler}. Hence, the QED magnetized vacuum gives rise also to a selective absorption of light depending on its polarization state~\footnote{Since we are supposing that $\omega \ll m_e$, $e^+e^-$ photo-production is kinematically forbidden (otherwise this process would dominate light absorption).}. Owing to the analogy with what happens in an optical medium, this phenomenon is called {\it dichroism} and is quantified by the absorption coefficients $a_{\parallel}$ and $a_{\bot}$ of the two propagating modes. Explicitly, one gets~\cite{Adler}
\begin{equation}
\label{am2}
a_{\parallel} = 0~,
\end{equation}
\begin{equation}
\label{am3}
a_{\bot} = 0.12 \left( \frac{B \, {\rm sin} \, \theta}{B_{\rm cr}} \right)^6 \, 
\left(\frac{\omega}{m_e} \right)^5 \, {\rm cm}^{- 1}~.
\end{equation} 
Notice that an unpolarized light beam becomes almost linearly polarized in the plane defined by the vectors $\bf B$ and $\bf k$, thanks to photon splitting.

So far, the polarization state of the monochromatic photon beam under consideration was supposed  arbitrary. Let us now investigate the consequences of birefringence and dichroism when the beam is {\it linearly} polarized at the beginning, at an angle $\varphi$ with respect to the plane defined by $\bf B$ and $\bf k$. We proceed schematically as follows.

Imagine first that only birefringence is operative. Then the two modes ${\gamma}_{\parallel}$ and ${\gamma}_{\bot}$ propagate with different speeds. Therefore, at any finite distance from the source, the beam polarization turns out to be {\it elliptical}. Even more explicitly, as the light beam moves along the $z$-axis, its electric field changes both direction and magnitude so as to trace a spiral (around the $z$-axis) with elliptical sections. After each $2 \pi$ rotation, a different projected ellipse gets singled out in the plane perpendicular to ${\bf k}$. Still -- as long as birefringence alone is concerned -- all such ellipses have {\it parallel} major axis, which is just an elementary manifestation of the composition of two harmonic motions along orthogonal directions. After travelling a distance $z$, the ellipticity is
\begin{equation}
\label{acd1}
{\epsilon}_{\rm QED} (z) = \frac{\omega z}{2} \left(n_{\parallel} - n_{\bot} \right) \, {\rm sin} \, 2 \varphi = \frac{\alpha}{60 \pi} \, \omega z \, \left(\frac{B \, \sin 
\theta}{B_{\rm cr}} \right)^2 \ {\rm  \sin \, 2 \varphi}~.
\end{equation}

Suppose next that only photon splitting is at work. Consequently, the resulting dichroism depletes the ${\gamma}_{\bot}$ mode, so that the electric field ${\bf E}_{\bot}$ in this mode gets reduced while the electric field ${\bf E}_{\parallel}$ in the other mode gets increased. Geometrically, the net result is a {\it rotation} of the electric field of the beam, namely of its polarization. After travelling a distance $z$, the rotation angle is
\begin{equation}
\label{acg1}
\Delta {\varphi}_{\rm QED} (z) = \frac{z}{4} \left(a_{\parallel} - a_{\bot} \right) \, {\rm sin} \, 2 \varphi = - \, 0.03 \, \left( \frac{B \, {\rm sin} \, {\theta}}{B_{\rm cr}} 
\right)^6 \, \left(\frac{\omega}{m_e} \right)^5 \, \left(\frac{z}{{\rm cm}} \right) \, {\rm  \sin \, 2 \varphi}~.
\end{equation}

On the whole, the light beam in question develops an {\it elliptical} polarization dictated by eq. (\ref{acd1}), with the ellipse's major axis {\it rotated} with respect to the initial polarization by an amount given by eq. (\ref{acg1}).           
  
In order to check the non-linear nature of QED, in 1979 Iacopini and Mimmo described an apparatus designed to measure the ellipticity induced by a strong magnetic field in a laser beam linearly polarized at the beginning~\cite{Iacopini}. Incidentally, they did not care about dichroism for a very good reason. According to the foregoing considerations, dichroism is {\it suppressed} by a factor ${\alpha}^{1/2} \, (B/B_{\rm cr})$ relative to birefringence in the QED magnetized vacuum, and so it is totally unobservable in laboratory experiments~\footnote{This statement was even more true in 1979. Hopefully, a considerable technological improvement might make dichroism observable in the future.}.

\section{Axions and Axion-Like Particles (ALPs)}

As is well known, non-perturbative effects produce the term $\theta g^2 F_a^{\mu \nu} \tilde F_{a \mu \nu} /32 {\pi}^2$ in the QCD effective lagrangian, where $\theta$ is an angle, while $g$ and 
$F_a^{\mu \nu}$ are the gauge coupling constant and the field strength of QCD ($\tilde F_{a \mu \nu}$ is the dual of $F_{a \mu \nu}$)~\cite{`tHooft}. All values of $\theta$ are allowed and theoretically on the same footing, but nonvanishing $\theta$ values produce a CP violation in the strong sector of the Standard Model. An additional source of
CP violation comes from the chiral transformation needed to bring the quark mass matrix ${\cal M}_q$ into diagonal form, and so the total strong CP violation is parametrized by 
$\bar{\theta} = \theta + \, {\rm Arg \ det} \, {\cal M}$. Observationally, a non-vanishing 
$\bar{\theta}$ would show up in a nonvanishing electric dipole moment for the neutron 
$d_n$~\cite{Baluni}. Consistency with the experimental upper bound $|d_n| < 3 \cdot 10^{- 26} \, 
{\rm e \, cm}$ requires $|\bar{\theta}| < 10^{- 9}$~\cite{neutrone}. 

A natural way out of this fine-tuning problem -- the ``strong CP problem''~\cite{SCPP} -- was proposed by Peccei and 
Quinn~\cite{Peccei}. Basically, the idea is to make the Standard Model lagrangian invariant under an additional {\it global} $U(1)_{\rm PQ}$ symmetry in such a way that the $\theta$-term can be rotated away. While this strategy can be successfully implemented, it turns out that the $U(1)_{\rm PQ}$ is spontaneously broken. Because of the Goldstone theorem, a physical spin-zero neutral boson is then present in the physical spectrum. Actually, things are slightly more complicated, because $U(1)_{\rm PQ}$ is also explicitly broken by the same non-perturbative effects which give rise to the $\theta$-term. Therefore, the would-be Goldstone boson becomes a pseudo-Goldstone boson with a nonvanishing mass, the standard {\it axion}~\cite{WW}. 

Qualitatively, the axion is quite similar to the pion, and it possesses Yukawa couplings to quarks which go like the inverse of the scale $f_a$ at which the $U(1)_{\rm PQ}$ is spontaneously broken. Moreover -- just like for the pion -- a two-photon coupling is generated at one-loop via the triangle graph (with internal fermion lines), which is described by the effective lagrangian
\begin{equation}
\label{aq5}
{\cal L}_{\phi \gamma} = - \frac{1}{4 M} \, F^{\mu \nu} \,  \tilde F_{\mu \nu} \, \phi = \frac{1}{M} \, {\bf E} \cdot {\bf B} \, \phi~,
\end{equation}
where $\phi$ denotes the axion field and the constant $M$ -- with the dimension of an energy -- is proportional to $f_a$. Notice that $M$ turns out to be independent of the mass of the fermion running in the loop.

In the original proposal~\cite{Peccei}, $U(1)_{\rm PQ}$ is spontaneously broken by two Higgs doublets which break $SU(2)$ X $U(1)$ spontaneously, so that $f_a$ coincides with the Fermi scale $G_F^{-1/2} \simeq 250 \, {\rm GeV}$. Correspondingly, the axion mass $m$ -- which scales like the inverse of $f_a$ -- turns out to be $m \sim 10^2 \, {\rm KeV}$. In addition, the axion is rather strongly coupled to quarks and induces {\it observable} nuclear de-excitation effects~\cite{Donnely}. In fact, it was soon realized that the original axion is experimentally ruled 
out~\cite{Zehnder}.

Remarkably enough, a slight change in perspective led to the resurrection of the axion strategy. It is easy to see how this comes about. Conflict with experiment arises because the original axion is too strongly coupled and too massive. But -- given the fact that both $m$ and all axion couplings go like the inverse of $f_a$ -- the axion becomes harmless provided one arranges $f_a \gg G_F^{-1/2}$. This is straightforwardly achieved by performing the spontaneous breakdown of $U(1)_{\rm PQ}$ with a Higgs field which is {\it neutral} under $SU(2)$ X $U(1)$. Clearly, this ``invisible axion'' scenario can be implemented naturally within a grand unified scheme~\cite{Glashow}. Indeed, this is a rather compelling option, because the $U(1)_{\rm PQ}$ symmetry is very unstable against a tiny perturbation -- even at the Planck scale -- unless it is protected by some discrete gauge symmetry~\cite{Roncadelli}. 

Several variations on this theme have been put forward~\cite{Dine,axion}. As a rule, the axion mass turns out to be model-independent and reads
\begin{equation}
\label{a6}
m \simeq 0.6 \left( \frac{10^7 \, {\rm GeV}}{f_a} \right) \, {\rm eV}~,
\end{equation}
while its two-photon inverse coupling constant in eq. (\ref{aq5}) is
\begin{equation}
\label{a7}
M \simeq 1.2 \cdot 10^{10} \, k \, \left( \frac{f_a}{10^7 \, {\rm GeV}} \right) \, \, {\rm GeV}~,
\end{equation}
where $k$ is a model-dependent parameter roughly of order one~\cite{cgn}. Hence, the axion is characterized by the following mass-coupling relation
\begin{equation}
\label{a8}
m \simeq 0.7 \cdot k \, \left( \frac{10^{10} \, {\rm GeV}}{M} \right) \, {\rm eV}~.
\end{equation}  

Cosmological considerations lead to a {\it lower} bound on the strenght of the photon-axion coupling~\cite{Kolb1990,Raffelt1990} in the form $f_a < 10^{13} \, {\rm GeV}$, which -- thanks to eqs. (\ref{a6}) and (\ref{a7}) -- entails in turn $M < 10^{16} \, {\rm GeV}$ and $m > 10^{ - 6} \, {\rm eV}$ (we have taken $k \simeq 1$). Depending on the actual value of $m$, the axion can be either an excellent cold dark matter candidate or a less appealing hot dark matter candidate, as discussed in the talk of Mirizzi at this Workshop~\cite{Mirizzi}.

Axion-Like Particles (ALPs) are quite similar to the axion and are described by the following effective lagrangian
\begin{equation}
\label{a5}
{\cal L}_{{\rm ALP}} = \frac{1}{2} \, \partial^{\mu} \phi \, \partial_{\mu} \phi - \frac{1}{2} 
\, m^2 \,{\phi}^2 - \frac{1}{4 M} \, F^{\mu \nu} \, \tilde F_{\mu \nu} \, \phi~,
\end{equation}
where $\phi$ now denotes the ALP field. Three points should be stessed. First, ALPs are supposed to be light, and for definiteness one assumes $m < 1 \, {\rm eV}$. Second, the mass $m$ and 
the inverse two-photon coupling $M$ of an ALP are regarded as {\it independent} parameters. Third, one just {\it assumes} the existence of a two-photon coupling as described by 
${\cal L}_{{\rm ALP}}$, without bothering about its origin. ALPs are predicted by many realistic extensions of the Standard Model and have attracted considerable interest in the last few years. Besides than in four-dimensional models~\cite{masso1}, they naturally arise in the context of compactified Kaluza-Klein theories~\cite{kk} as well as in superstring theories~\cite{superstring}. In the rest of this Proceeding, we shall be concerned more generally with ALPs~\footnote{We stress that the considerations to follow also apply to a {\it scalar} light boson (with somewhat trivial modifications), provided $F^{\mu \nu} \, \tilde F_{\mu \nu} \, \phi$ is replaced by $F^{\mu \nu} \, F_{\mu \nu} \, \phi$ in eq. (\ref{a5}). In the latter case, one has however to worry about possible violations of the equivalence principle~\cite{carlo}.}.

Astrophysics provides a stark {\it upper} bound on the strenght of the photon coupling for 
{\it any} ALP. The argument is as follows~\cite{Raffelt1990}. Because of its two-photon coupling, an ALP can be photo-produced -- via Primakoff scattering -- in the inner region of main-sequence and red-giant stars, where temperature is high and matter is fully ionized. More explicitly, a thermal photon can exchange a virtual photon with an ion and become an ALP. Because its mean-free-path is much larger than the stellar radius, the ALP escapes, thereby carrying off energy. Owing to the virial equilibrium, the core has a negative specific heat. Therefore it reacts to such an energy loss by getting hotter. As a result, the rate of nuclear reactions sharply increases, thereby changing the observed properties of stars. Yet, current models of stellar evolution are in fairly good agreement with observations. Hence, the two-photon inverse coupling constant $M$ has to be large enough to provide a sufficient suppression of unwanted ALP effects. This argument has been applied in a quantitative fashion to stars of different kinds, with the result~\cite{Raffelt1990} 
\begin{equation}
\label{Zak5}
M > 10^{10} - 10^{11} \, {\rm GeV}~. 
\end{equation}
Quite remarkably, just the same conclusion is reached by the CAST experiment at CERN, designed 
to detect ALPs (axions) emitted by the Sun~\cite{cast} (more about this, later).

\section{Photon-ALP mixing}

Consider the $\gamma \gamma \phi$ mixing term in ${\cal L}_{{\rm ALP}}$
\begin{equation}
\label{ak5}
 - \frac{1}{4 M} \, F^{\mu \nu} \, \tilde F_{\mu \nu} \, \phi = \frac{1}{M} \, {\bf E} \cdot {\bf B} \, \phi~.
\end{equation}
In the presence of a magnetic field ${\bf B}$, one $\gamma$ has to be replaced by a ``magnetic-field photon'' ${\gamma}_B$, thereby implying that the propagation eigenstates {\it differ} from the corresponding interaction eigenstates. Hence, photon-ALP {\it interconversion} takes place. Notice that ${\bf B}$ does not play the role of an energy source but merely acts like a catalyst.

Below, we address two aspects of this phenomenon, which are of paramount importance for the PVLAS experiment and its competitors.

\subsection{Photon-ALP oscillations}

{\it Coherent} mixing is a concept quite familiar to the audience of this Workshop, and so no additional comment is necessary: just think of coherent flavour mixing in the case of massive neutrinos. Indeed, coherent photon-ALP interconversion is similar in nature to neutrino 
oscillations~\cite{neutrino oscil}, apart from the fact that here the magnetic field ${\bf B}$ is needed in order to compensate for the spin mismatch. 

Let us now look at photon-ALP oscillations again in the case of a monochromatic photon beam with frequency $\omega$ travelling along the $z$-direction, assuming that the magnetic field 
${\bf B}$ is {\it arbitrary}. 

Clearly, the beam propagation is described by the {\it second-order} coupled Klein-Gordon and Maxwell equations dictated by lagrangian ${\cal L}_{{\rm HEW}} + {\cal L}_{{\rm ALP}}$. Quite often -- due to the very small ALP mass -- one is interested in the regime in which 
$\omega \gg m$. In such a situation, the short-wavelength approximation (WKB approximation) can be successfully applied and turns the above wave equation into the {\it first-order} propagation equation~\cite{RaffeltStodolsky,Lai}
\begin{equation}
\label{aa6}
i \, \frac{d}{d z} | \psi (z) \rangle = {\cal M} | \psi (z) \rangle~,
\end{equation}
with
\begin{equation}
\label{aa7}
| \psi (z) \rangle \equiv A_x (z) \, |x \rangle + A_y (z) \, |y \rangle + {\phi} (z) \, 
| \phi \rangle~,        
\end{equation} 
where $| x \rangle$ and $| y \rangle$ are the two photon linear polarization states along the 
$x$ and $y$ axis, respectively, and $| \phi \rangle$ denotes the ALP state. Hence, we see that in this approximation the behaviour of a relativistic photon/ALP beam becomes equal to that of a 
non-relativistic three-level system.

In the $\{| x \rangle, | y \rangle, | \phi \rangle \}$ basis, the mixing matrix ${\cal M}$ has the general form
\begin{equation}
\label{aa8}
{\cal M} = \left(
\begin{array}{ccc}
\omega + \Delta_{xx} & \Delta_{xy} & B_x /2M \\
\Delta_{yx} & \omega + \Delta_{yy} & B_y /2M \\
B_x /2M & B_y /2M & \omega - m^2 /2 \omega \\
\end{array}
\right)~.
\end{equation}
While the terms appearing in the third row and column of ${\cal M}$ have an evident physical meaning, the $\Delta$-terms require some explanation. Generally speaking, they reflect the properties of the medium in which the beam propagates~\footnote{These properties are not described by ${\cal L}_{{\rm HEW}} + {\cal L}_{{\rm ALP}}$ and can be accounted for by inserting in such a lagrangian a suitable dielectric tensor~\cite{Yoshimura}} as well as the QED vacuum effects addressed in Section 2. Off-diagonal $\Delta$-terms directly mix the photon polarization states and typically give rise to Faraday rotation. 

When the magnetic field is {\it homogeneous}, we can choose the $y$-axis along the 
projection of ${\bf B}$ perpendicular to the $z$-axis. Correspondingly we have $B_x = 0$, $B_y = B \, {\rm sin} \, {\theta}$, $x = {\bot}$, $y = {\parallel}$, and the mixing matrix takes the form
\begin{equation}
\label{a9}
{\cal M} = \left(
\begin{array}{ccc}
\omega + \Delta_{\bot} & \Delta_R & 0 \\
\Delta_R & \omega + \Delta_{\parallel} & B \, {\rm sin} \, {\theta} /2M \\
0 & B \, {\rm sin} \, {\theta} /2M & \omega - m^2 /2 \omega \\
\end{array}
\right)~.
\end{equation}
In general, the diagonal $\Delta$-terms receive three different contributions, and so we write
\begin{equation}
\label{a10}
{\Delta}_{\parallel, \bot} = {\Delta}_{\parallel, \bot}^{\rm QED} + {\Delta}_{\parallel, 
\bot}^{\rm PL} + {\Delta}_{\parallel, \bot}^{\rm CM}~.
\end{equation}
The terms ${\Delta}_{\parallel, \bot}^{\rm QED}$ represent the QED vacuum magnetic effects. They follow directly from eqs. (\ref{a2}), (\ref{a3}) and read
\begin{equation}
\label{a11}
{\Delta}_{\parallel}^{\rm QED} = \left( n_{\parallel} - 1 \right) \omega = \frac{7}{2} \, \left( \frac{\alpha}{45 \pi} \right) \, \left( \frac{B \, {\rm sin} \, \theta}{B_{\rm cr}} 
\right)^2 \omega~,
\end{equation}
\begin{equation}
\label{a12}
{\Delta}_{\bot}^{\rm QED} = \left( n_{\bot} - 1 \right) \omega = \frac{4}{2} \, \left( \frac{\alpha}{45 \pi} \right) \, \left( \frac{B \, {\rm sin} \, \theta}{B_{\rm cr}} 
\right)^2 \omega~. 
\end{equation}
In the presence of a plasma, charge screening effects produce an effective photon mass given by the plasma frequency $\omega_{\rm pl}^2 = 4 \pi \alpha \, n_e/ m_e$ -- here $n_e$ denotes the electron density -- and the resulting contribution is 
\begin{equation}
\label{a13}
{\Delta}^{\rm PL}_{\parallel, \bot} = - \, \frac{\omega_{\rm pl}^2}{2 \omega} = - \, 
\frac{2 \pi \alpha \, n_e}{m_e \, \omega}~.
\end{equation}
Furthermore, the term ${\Delta}_{\parallel, \bot}^{\rm CM}$ describes the Cotton-Mouton effect, namely birefringence in a fluid due to a transverse magnetic field. Finally, the term 
$\Delta_R$ accounts for Faraday rotation. 

Restricting the attention to the case in which Cotton-Mouton and Faraday effects are unimportant, both ${\Delta}_{\parallel, \bot}^{\rm CM}$ and $\Delta_R$ can be discarded. Consequently, the longitudinal component of the magnetic field $B \, {\rm cos} \, {\theta}$ disappears from ${\cal M}$, ${\gamma}_{\bot}$ decouples away and {\it only} ${\gamma}_{\parallel}$ mixes 
with $\phi$~\footnote{Since environmental mixing effects are neglected, the letter circumstance directly follows from ${\cal L}_{\phi \gamma}$ also for an arbitrary magnetic field.}. As a result, the mixing matrix ${\cal M}$ reduces to the two-dimensional form
\begin{equation}
\label{a14}
{\cal M}_0 = \left(
\begin{array}{cc}
\omega + \Delta_{\parallel} & B \, {\rm sin} \, {\theta} /2M \\
B \, {\rm sin} \, {\theta} /2M & \omega - m^2 /2 \omega \\
\end{array}
\right)~.
\end{equation}
This matrix can be diagonalized by an orthogonal transformation with rotation angle
\begin{equation}
\label{a15}
\Theta = \frac{1}{2} \, {\rm arctg} \left( \frac{B \, {\rm sin} \, \theta /M}{\Delta_{\parallel} + m^2/2 \omega} \right)
\end{equation}
and -- in complete analogy with the case of neutrino oscillations~\cite{neutrino oscil} -- the probability that a ${\gamma}_{\parallel}$ photon will be converted into an ALP after a distance 
$z$ is 
\begin{equation}
\label{a16}
P ({\gamma}_{\parallel} \to {\phi}) = {\rm sin}^2 2 \Theta \  {\rm sin}^2 
\left( \frac{\Delta_{\rm osc} \, z}{2} \right) = \left( \frac{B \, {\rm sin} \, \theta }{M \, {\Delta}_{\rm osc}} \right)^2 \  {\rm sin}^2 
\left( \frac{\Delta_{\rm osc} \, z}{2} \right)~,
\end{equation}
where the oscillation wavenumber is
\begin{equation}
\label{a17}
{\Delta}^2_{\rm osc} = \left( \Delta_{\parallel} + \frac{m^2}{2 \omega} \right)^2 + 
\left( \frac{B \, {\rm sin} \, \theta}{M} \right)^2~,
\end{equation}
so that the  oscillation length is just $L_{\rm osc} = 2 \pi / {\Delta}_{\rm osc}$. 

\subsection{Polarization effects}

A further consequence of the mixing term (\ref{ak5}) concerns the polarization state of the 
photon beam in question. As realized by Maiani, Petronzio and Mimmo~\cite{Maiani}, this comes about in two distinct ways. 

We have seen in Section 2 that the exchange of virtual fermions yields a non-trivial contribution to the photon propagator in the presence of a magnetic field (Delbr\"uck scattering), which affects the velocity of light and gives rise to vacuum birefringence. A further contribution of the same kind arises from the exchange of {\it virtual} ALPs, namely from the diagram in which two $\gamma \gamma \phi$ vertices are joined together by the $\phi$ line and two external photon lines are actually $\gamma_B$ photons. Again, the corresponding amplitude depends on the polarization of the incoming photon, and so it is an {\it additional} source of vacuum {\it birefringence}. As before, the photon propagation eigenstates are {\it linear} polarization modes -- denoted by 
${\gamma}^{\prime}_{\parallel}$ and ${\gamma}^{\prime}_{\bot}$ --  but here a small complication arises since the mass eigenstates differ from the interaction eigenstates. This point can be clarified by supposing (for simplicity) that the magnetic field is {\it homogeneous}. Then we know that ${\gamma}_{\bot}$ decouples from $\phi$, thereby implying that it remains a propagation eigenstate with refractive index still given by eq. (\ref{a3}). That is, we have
${\gamma}^{\prime}_{\bot} = {\gamma}_{\bot}$ and 
\begin{equation}
\label{aq18}
n^\prime_{\bot} = n_{\bot}~. 
\end{equation}
On the other hand, ${\gamma}_{\parallel}'$ is that particular linear combination of 
${\gamma}_{\parallel}$ and $\phi$ which diagonalizes the mixing matrix ${\cal M}_0$ in eq. (\ref{a14}). A straightforward calculation shows that eq. (\ref{a2}) gets presently replaced by
\begin{equation}
\label{a18}
n^\prime_{\parallel} = n_{\parallel} + \frac{1}{2 \omega} \left\{ 
\left[ \left( \frac{B \, {\rm sin} \, \theta}{M} \right)^2 + \left( {\Delta}_{\parallel} + 
\frac{m^2}{2 \omega} \right)^2 \right]^{1/2} - \left( {\Delta}_{\parallel} + 
\frac{m^2}{2 \omega}   \right) \right\}~. 
\end{equation}

Next, we address the effect arising from the production of {\it real} ALPs, occurring via the $\gamma \gamma \phi$ vertex with one photon line representing a $\gamma_B$ photon. Because only 
${\gamma}_{\parallel}$ mixes with ${\phi}$, photon-ALP conversion is obviously an {\it additional} source of vacuum {\it dichroism}. Also in this case dischroism is quantified by the absorption coefficients $a^{\prime}_{\parallel}$ and $a^{\prime}_{\bot}$ of the two propagating modes.

Suppose now that the light beam under consideration is {\it linearly} polarized at the 
beginning, at an angle $\varphi$ with respect to the plane defined by $\bf B$ and $\bf k$. How is its polarization state affected by ALPs? 

We know that birefringence gives rise to an {\it elliptical} polarization, whereas dichroism produces a {\it rotation} of the ellipse's major with respect to the initial polarization. Manifestly, these effects are qualitatively {\it identical} to those arising from the QED magnetized vacuum alone. Quantitatively, Maiani, Petronzio and 
Mimmo~\cite{Maiani,RaffeltStodolsky} have computed the induced ellipticity 
$\epsilon_{\rm ALP} (z)$ and rotation angle $\Delta \varphi_{\rm ALP} (z)$. In the approximation -- appropriate for the PVLAS experiment -- in which the oscillation wavenumber is dominated by the ALP mass term~\footnote{See eq. (\ref{a17}).}, they find 
\begin{equation}
\label{aWE18}
{\epsilon}_{\rm ALP} (z) = \frac{1}{2} \, \left( \frac{B \, {\rm sin} \, \theta \, \omega}{m^2 \, M} \right)^2 \, \left[ \frac{m^2 z}{2 \omega} - {\rm sin} \left( \frac{m^2 z}{2 \omega} \right) \right] \, {\rm sin} \, 2 \varphi~, 
\end{equation}
and
\begin{equation}
\label{aWp18}
\Delta {\varphi}_{\rm ALP} (z) = \left( \frac{B \, {\rm sin} \, \theta \, \omega}{m^2 \, M} \right)^2 \, 
{\rm sin}^2 \left( \frac{m^2 z}{4 \omega} \right) \, {\rm sin} \, 2 \varphi~. 
\end{equation}

Finally, it has been recently realized that the existence of ALPs makes photon splitting highly enhanched by inhomogeneities of the magnetic field similar to those expected to show up in the PVLAS experiment~\cite{Giovannini}.

\section{PVLAS and beyond}

Over the last two decades, various proposals have been put forward to detect axions through their two-photon coupling in non-accelerator situations. Generic ALPs can also be discovered by these techniques, provided of course that their mass $m$ and inverse two-photon coupling constant $M$ fall into suitable ranges dictated by the experimental setup. 

A strategy addresses the possibility of converting axions present in the laboratory into photons. As suggested by Sikivie~\cite{Sikivie}, Galactic dark matter axions are expected to excite a proper TM mode of a {\it tunable} microwave cavity -- permeated by a strong magnetic field -- when resonance takes place, namely when the characteristic frequency of a cavity TM proper mode 
happens to coincide with the axion mass. So far, no positive signal has been reported by experiments of this kind~\cite{Bradley}. Although this method can also be applied to axions coming from the Sun, they can be more easily detected by back-conversion into $X$-ray photons in a ``magnetic telescope''~\cite{vanbibber}. Recently, the latter technique has been implemented by the above-mentioned CAST experiment at CERN, leading to the lower bound $M > 1.14 \cdot 10^{10} \, {\rm GeV}$ for $m < 0.02 \, {\rm eV}$~\cite{cast}.

An alternative strategy relies upon axion effects produced on photon propagation in the presence of a strong magnetic field. As we have seen, an initially {\it linearly}-polarized light beam is expected to become {\it elliptically} polarized, with the ellipse's major axis {\it rotated} with respect to the initial polarization. Because of eqs. (\ref{acd1}) and (\ref{aWE18}), the total ellipticity is
\begin{equation}
\label{sacd1}
\epsilon (z) = \epsilon_{\rm QED} (z) + \epsilon_{\rm ALP} (z)~.
\end{equation}
Similarly, by eqs. (\ref{acg1}) and (\ref{aWp18}) the total rotation angle reads
\begin{equation}
\label{axWp18}
\Delta \varphi (z) = \Delta {\varphi}_{\rm QED} (z)  + \Delta {\varphi}_{\rm ALP} (z)~. 
\end{equation}
The crucial point -- noted by Maiani, Petronzio and Mimmo~\cite{Maiani} -- is that the  contribution in question {\it dominates} over the QED one in a physically interesting region of the $m-M$ parameter plane for the axion~\footnote{No observable rotation is expected to arise in the QED vacuum because dichroism is suppressed with respect to birefringence, as pointed out in Section 2.}. But then -- thanks to eqs. (\ref{aWE18}) and (\ref{aWp18}) -- both $m$ and $M$ are {\it uniquely determined} once $\epsilon (z)$ and $\Delta \varphi (z)$ are {\it measured}, which is the ultimate experimental goal. Needless to say, also ALPs can be searched for in this way. Notice that here it is completely irrelevant whether ALPs are present or not in the laboratory.

Using an experimental setup based on the latter method, in 1992 Mimmo and collaborators established the lower bound $M > 2.8 \cdot 10^{6} \, {\rm GeV}$ for $m < 10^{ - 3} \, {\rm eV}$~\cite{Brookhaven}. In 2005 -- by exploiting a similar but more sophisticated technique -- the PVLAS collaboration has reported positive evidence for an anomalously large value of the rotation angle $\Delta \varphi$ in an initially linearly-polarized laser beam undergoing multiple reflection in a $5 \, T$  magnetic field~\cite{pvlas1,pvlas2}. In addition, also the beam ellipticity $\epsilon$ has been determined. {\it Assuming} that the effect is indeed brought about by an ALP, the corresponding physical parameters turn out to lie in the range $ 1.0 \cdot 10^{- 3} \, {\rm eV} \leq m \leq 1.5 \cdot 10^{- 3} \, {\rm eV} $ and $2 \cdot 10^{5} \, {\rm GeV} \leq M\leq 6 \cdot 10^{5} \, {\rm GeV}$. 

Manifestly, a look back at eq. (\ref{a8}) shows that the ALP in question {\it cannot} be the axion. Moreover, the quoted value of $M$ {\it violates} the astrophysical bound 
(\ref{Zak5}) by about five orders of magnitudes. The only way out of this difficulty is evidently to suppose that ALPs are {\it not} emitted by stars. Conceivably, high-temperature effects~\cite{Mohapatra} or plasma effects~\cite{Massoprl} typical of stellar interiors can suppress their production. Alternatively, ALPs might still be produced but should remain confined inside the inner region of stars in a manner consistent with the observed properties~\cite{masso2005}. In either case, {\it new physics} at energy as low as a few ${\rm KeV}$ is required. This issue is discussed in the talk of Ringwald at this Workshop~\cite{Ringwald}. 

Without any doubt, the need for {\it independent} tests of the PVLAS claim looks compelling. 

Concerning the latter point, a few options are presently considered and some experiments will soon start. 

In the first place, an experiment similar to PVLAS should be performed, however with a different magnet and a laser beam with a different frequency. This is the case e. g. for the BMV project at LNCMR~\cite{askenazy}.

A somewhat different method exploits the idea of {\it photon regeneration}, which can be illustrated as follows. Suppose that a light beam is shone across a magnetic field, so that some photons are converted into ALPs. If a screen is put on the beam path, photons are completely absorbed, but ALPs are not. Hence, ALPs emerge on the other side of the screen and undergo photon back-conversion if a second magnetic field is present. Detection of photons on the latter side of the screen would then be an unambiguous signal of photon-ALP conversion~\cite{photreg}. Several experiments of this kind are planned, either with an ordinary laser beam or with a synchrotron 
$X$-ray beam produced in a free-electron laser~\cite{Ringwald}. 

\section{PVLAS in the sky}

Remarkably enought, some astrophysical settings are characterized by environmental conditions quite similar to those occurring in a laboratory experiment (low temperature and low electron density). In such a  situation, the photon-ALP mixing -- with the {\it same} strenght as that claimed by PVLAS -- should be at work. Below, we address three cross-check based on this strategy.

A possibility is offered by the recently discovered double pulsar system J0737-3039, which has an orbital period of 2.45 hours and is seen almost edg-on~\cite{DoublePulsar}. Consider the gamma-ray beam emitted by one component pulsar, say pulsar $A$. When $A$ lies almost exactly behind 
$B$, the beam experiences the strong magnetic field produced by pulsar $B$, in which photon-ALP oscillations should occur. As a result, a characteristic attenuation pattern of the beam is expected. Assuming realistic values for the parameters involved, it turns out that the upcoming GLAST mission can check the PVLAS claim. Even the no-detection of an attenuation at the $10 \, \%$ level would be inconsistent with the PVLAS claim, as shown in the exclusion plot in Fig. 1~\cite{drrb}. 
\begin{figure}[h]
\begin{center}
\includegraphics[width=8cm]{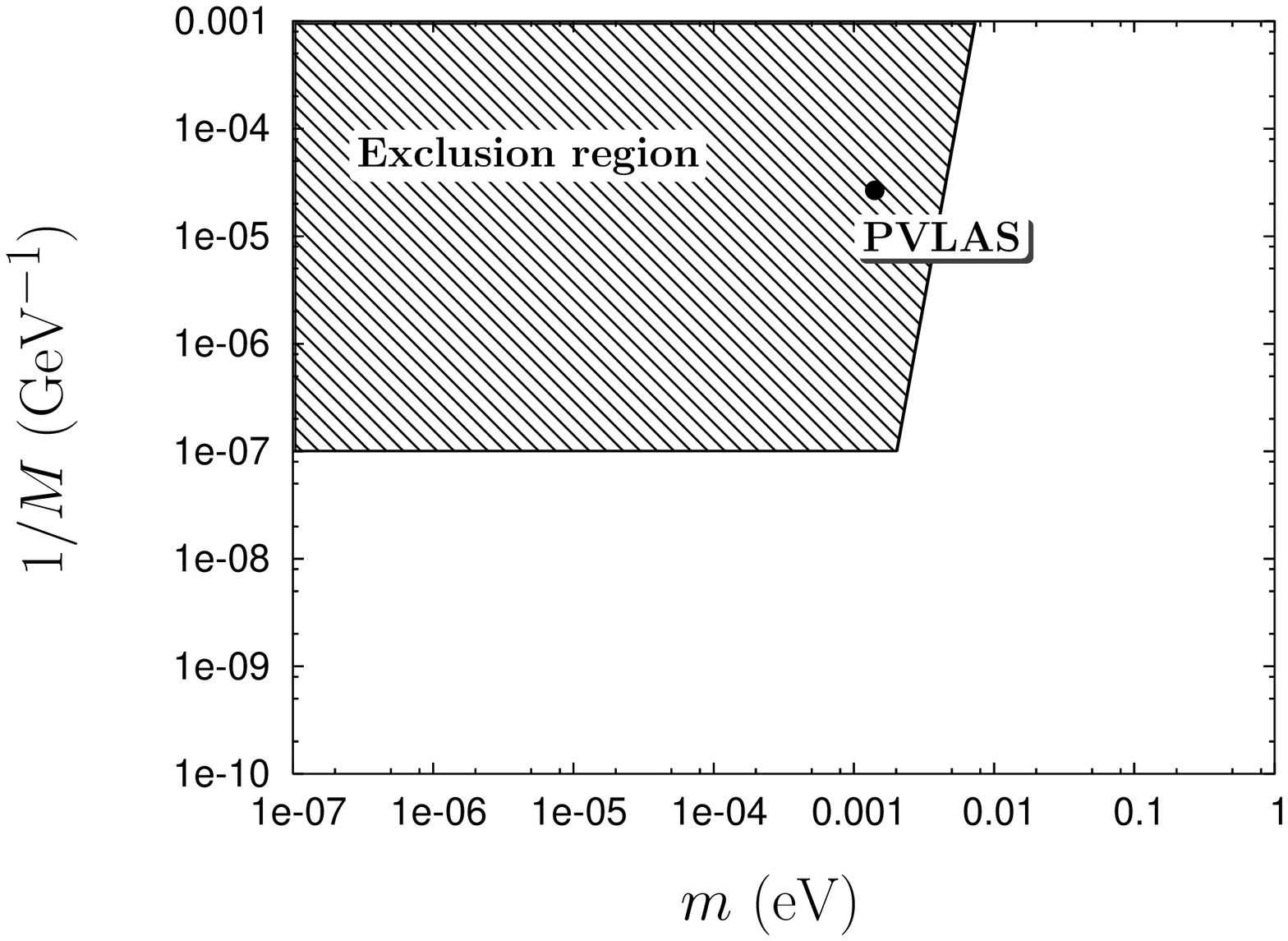}
%\caption{\label{fig3}{Exclusion region in the case that the existence of the attenuation is 
%excluded at $10\,\%$ level.}}
\end{center}
\end{figure}

A different proposal concerns the peculiar dimming of an extragalactic source arising from 
photon-ALP oscillations induced by the turbolent magnetic field in the Milky Way. This effect is expected to show up for photons with energy larger than $10 \, {\rm TeV}$, and so could be detected with Imaging Atmospheric Cherenkov Telescopes~\cite{MRS}.

Finally, a very interesting astrophysical realization of the photon-regeneration strategy concerns the Sun. Suppose that a gamma-ray source (distant quasar) is occulted by the Sun. Ordinarily, one would expect the source to become invisible. However -- if the PVLAS claim is correct -- this would not necessarily be the case and the source might still be seen. For, some of the photons approaching the far side of the Sun would be converted to ALPs in the solar magnetic field. These ALPs would then traverse the Sun unimpeded and would next undergo partial back-conversion into gamma-ray photons in the magnetic field on the near side of the Sun. It has been claimed that GLAST could see this effect~\cite{Fairbairn}.

\section{Acknowledgments}

I thank Nanni Bignami, Gianni Carugno, Claudio Corian\`o, Arnaud Dupays, Maurizio Lusignoli, Luciano Maiani, Alessandro Mirizzi, Antonello Polosa, Carlo Rizzo and the members of the PVLAS collaboration for discussions about the topics reported here. Countless discussions with Mimmo Zavattini remain unforgettable for me. I am also very grateful to Milla for giving me the opportunity to review at this stimulating Workshop part of the scientific activity of a great physicist and a dear friend.

\end{document}